\documentclass[a4paper]{spie}  
\usepackage[]{graphicx}

\def\spose#1{\hbox to 0pt{#1\hss}}
\def\simlt{\mathrel{\spose{\lower 3pt\hbox{$\mathchar"218$}}
     \raise 2.0pt\hbox{$\mathchar"13C$}}}
\def\simgt{\mathrel{\spose{\lower 3pt\hbox{$\mathchar"218$}}
     \raise 2.0pt\hbox{$\mathchar"13E$}}}

\title{MICADO: first light imager for the E-ELT} 

\author[a]{R.~Davies}
\author[a]{J.~Schubert}
\author[a]{M.~Hartl}
\author[b]{J.~Alves}
\author[c]{Y.~Cl\'enet}
\author[d]{F.~Lang-Bardl}
\author[e]{H.~Nicklas}
\author[f]{J.-U.~Pott}
\author[g]{R.~Ragazzoni}
\author[h]{E.~Tolstoy}
\author[i]{T.~Agocs}
\author[e]{H.~Anwand-Heerwart}
\author[f]{S.~Barboza}
\author[c]{P.~Baudoz}
\author[a,d]{R.~Bender}
\author[f]{P.~Bizenberger}
\author[c]{A.~Boccaletti}
\author[i]{W.~Boland}
\author[j]{P.~Bonifacio}
\author[f]{F.~Briegel}
\author[c]{T.~Buey}
\author[c]{F.~Chapron}
\author[c]{M.~Cohen}
\author[b]{O.~Czoske}
\author[e]{S.~Dreizler}
\author[g]{R.~Falomo}
\author[k]{P.~Feautrier}
\author[a]{N.~F\"orster~Schreiber}
\author[c]{E.~Gendron}
\author[a]{R.~Genzel}
\author[f]{M.~Gl\"uck}
\author[c]{D.~Gratadour}
\author[l]{R.~Greimel}
\author[a,d]{F.~Grupp}
\author[d]{M.~H\"auser}
\author[a]{M.~Haug}
\author[f]{J.~Hennawi}
\author[d]{H.-J.~Hess}
\author[a]{V.~H\"ormann}
\author[f]{R.~Hofferbert}
\author[a,d]{U.~Hopp}
\author[c]{Z.~Hubert}
\author[m]{D.~Ives}
\author[b,n]{W.~Kausch}
\author[m]{F.~Kerber}
\author[d]{H.~Kravcar}
\author[o]{K.~Kuijken}
\author[d]{F.~Lang-Bardl}
\author[l]{M.~Leitzinger}
\author[b]{K.~Leschinski}
\author[h]{D.~Massari}
\author[j]{S.~Mei}
\author[c]{F.~Merlin}
\author[f]{L.~Mohr}
\author[d]{A.~Monna}
\author[f]{F.~M\"uller}
\author[i]{R.~Navarro}
\author[a]{M.~Plattner}
\author[n]{N.~Przybilla}
\author[p]{R.~Ramlau}
\author[m]{S.~Ramsay}
\author[l]{T.~Ratzka}
\author[e]{P.~Rhode}
\author[d]{J.~Richter}
\author[f]{H.-W.~Rix}
\author[f]{G.~Rodeghiero}
\author[f]{R.-R.~Rohloff}
\author[c]{G.~Rousset}
\author[a]{R.~Ruddenklau}
\author[n]{V.~Schaffenroth}
\author[d]{J.~Schlichter}
\author[c]{A.~Sevin}
\author[o]{R.~Stuik}
\author[a]{E.~Sturm}
\author[a,d]{J.~Thomas}
\author[i]{N.~Tromp}
\author[g]{M.~Turatto}
\author[h]{G.~Verdoes-Kleijn}
\author[c]{F.~Vidal}
\author[q]{R.~Wagner}
\author[d]{M.~Wegner}
\author[b]{W.~Zeilinger}
\author[b]{B.~Ziegler}
\author[k]{G.~Zins}
\affil[a]{Max Planck Institute for extraterrestrial Physics, 
85748 Garching, Germany}
\affil[b]{Institute for Astrophysics, University of Vienna, T\"urkenschanzstrasse 17, 1180 Wien, Austria}
\affil[c]{LESIA, Observatoire de Paris, PSL Research University, CNRS, France}
\affil[d]{University Observatory of Munich, Scheinerstrasse 1, 81679 M\"unchen, Germany}
\affil[e]{Institute for Astrophysics, Friedrich-Hund Platz 1, 37077 G\"ottingen, Germany}
\affil[f]{Max Planck Institute for Astronomy, K\"onigstuhl 17, 69117 Heidelberg, Germany}
\affil[g]{INAF - Astronomical Observatory of Padova, Vicolo dell'Osservatorio 5, 35122, Padova, Italy}
\affil[h]{Kapteyn Astronomical Institute, 
Postbus 800, 9700 AV Groningen, The Netherlands}
\affil[i]{NOVA-ASTRON, Oude Hoogeveensedijk, 7991 PD, Dwingeloo, The Netherlands}
\affil[j]{GEPI, Observatoire de Paris, PSL Research University, CNRS, France}
\affil[k]{IPAG, Universit\'e Grenoble Alpes, CNRS, 38041 Grenoble, France}
\affil[l]{Institute of Physics, Karl-Franzens-Universit\"at, Universit\"atsplatz 5/II, 8010 Graz, Austria}
\affil[m]{ESO, Karl-Schwarzschild-Strasse 2, 85748, Garching, Germany}
\affil[n]{Institute for Astro and Particle Physics, 
Technikerstrasse 25/8, 6020 Innsbruck, Austria}
\affil[o]{Leiden Observatory, University of Leiden, Postbus 9513, 2300 RA Leiden, The Netherlands}
\affil[p]{Industrial Mathematics Institute, 
Altenbergerstrasse 69, 4040 Linz, Austria}
\affil[q]{Johann Radon Institute, 
Altenbergerstrasse 69, 4040 Linz, Austria}

\authorinfo{Further author information: send correspondence to R. Davies, email davies@mpe.mpg.de}
 
\begin{document} 
  \maketitle 

\begin{abstract}
MICADO will equip the E-ELT with a first light capability for diffraction limited imaging at near-infrared wavelengths. The instrument's observing modes focus on various flavours of imaging, including astrometric, high contrast, and time resolved. There is also a single object spectroscopic mode optimised for wavelength coverage at moderately high resolution. This contribution provides an overview of the key functionality of the instrument, outlining the scientific rationale for its observing modes. The interface between MICADO and the adaptive optics system MAORY that feeds it is summarised.
The design of the instrument is discussed, focussing on the optics and mechanisms inside the cryostat, together with a brief overview of the other key sub-systems.
\end{abstract}


\keywords{ELT, near-infrared, adaptive optics, camera}

\section{INTRODUCTION}
\label{sec:intro}  

MICADO is the Multi-AO Imaging Camera for Deep Observations, that is being designed and built by a consortium of institutes in Germany, France, Netherlands, Austria, and Italy. It will equip the European Extremely Large Telescope (E-ELT) with a first light capability for diffraction limited imaging at near-infrared wavelengths (0.8--2.4\,$\mu$m). The instrument is optimised to work with the laser guide star multi-conjugate adaptive optics (MCAO) module developed by the MAORY consortium\cite{dio16}. It also includes a jointly developed single-conjugate adaptive optics (SCAO) mode\cite{cle16} that uses just a single natural guide star. MICADO has to be able to function in a 'stand-alone' deployment configuration that involves operating the instrument with the SCAO wavefront sensor but without the full MCAO bench. It is required for the acceptance, integration and verification (AIV) phase of the MICADO project, and is also part of the programmatic risk mitigation strategy for the E-ELT.

Following the approval of the E-ELT construction proposal in December~2014, the Agreement to design and build MICADO was signed in September~2015, and the Phase~B Kick-Off meeting was held shortly afterwards.
The schedule for MICADO is tied to that of the E-ELT, for which first light is planned to be in 2024.
In order to achieve this, and yet also to accommodate development of the interfaces to the observatory (and also to MAORY) that is expected to occur during the early stages of the project, the preliminary design phase (Phase~B) of MICADO has been scheduled to last for 3 years, with the associated design review expect to be held towards the endof 2018.
Mid-way through that phase, there will be an internal review of the system and sub-system specifications and interfaces.
While the design has evolved considerably since the end of Phase~A\cite{dav10} in 2009, the rationale for the instrument remains fundamentally the same.

The key capabilities of MICADO exploit the most unique features of the E-ELT: sensitivity and resolution, precision astrometry, and wide wavelength coverage spectroscopy.
The primary observing mode is imaging, with a focus on astrometry\cite{tri10,fri10}. To achieve the stability necessary to provide spatial resolution better than 10\,mas and astrometric precision below 50\,$\mu$as, the instrument is supported above the Nasmyth platform in a gravity invariant orientation, includes an optical path comprising entirely of fixed mirrors, uses a state-of-the-art atmospheric dispersion corrector, and has a dedicated astrometric calibration plan and data pipeline. The array of $3\times3$ 4k$^2$ detectors at the focal plane can image a small field of about 20$^{\prime\prime}$ with a fine 1.5\,mas pixel sampling that is especially useful in very crowded fields or at short wavelengths; as well as a large field more than 50$^{\prime\prime}$ across, with a coarser 4\,mas pixel scale that still fully samples the H- and K-band diffraction limit.
In both cases, a wide selection of broad and narrow band filters are available. This mode will provide comparable sensitivity to the James Webb Space Telescope at 6 times better spatial resolution, and enable proper motions as small as 5\,km\,s$^{-1}$ to be measured at distances of up to 100\,kpc.
While exoplanets are among the primary science drivers for the E-ELT, a dedicated camera for such studies will not be available for a number of years after first light. As such, a secondary role of MICADO is to bridge this gap, using coronagraphy to providing a high contrast imaging capability.
This is enabled via a variety of focal plane and/or pupil plane masks, and is envisaged to make use of angular differential imaging techniques. Its novel feature will be the very small angular scales on which it will be possible to detect exoplanets. 
A slit spectroscopic mode is also included in MICADO, optimised for compact objects. It emphasizes simultaneous wavelength coverage at moderately high resolution: covering IJ and HK bands together at a spectral resolution of $R\sim15000$ for point sources (the resolution will be lower for extended objects that fill the width of the slit). Slit widths suitable for compact and extended objects will be provided.
Time resolved imaging is enabled by defining suitable windows on the detector which enable the frame rate to be in the range 1--100\,Hz, and providing precise ($<$1\,ms time) stamping of the frames.

\section{Science Drivers and Key Capabilities}
\label{sec:science}

MICADO has the potential to address a large number of science topics that span the key elements of modern astrophysics, and has clear synergies with other facilities and instruments such as ALMA, HARMONI\cite{tha16}, and METIS\cite{bra16}.
The science drivers for MICADO cover six main themes: (i) galaxy evolution at high redshift, (ii) black holes in galaxy centres, including the Galactic Center, (iii) resolved stellar populations, including photometry in galaxy nuclei, the initial mass function in young star clusters, and intermediate mass black holes in globular clusters, (iv) characterisation of exoplanets and circumnuclear disks at small angular scales, (v) the solar system, and (vi) time resolved phenomena around neutron stars and stellar mass black holes.

\subsection{Galaxy Evolution at high redshift}

\begin{figure}
\begin{center}
\includegraphics[width=12cm]{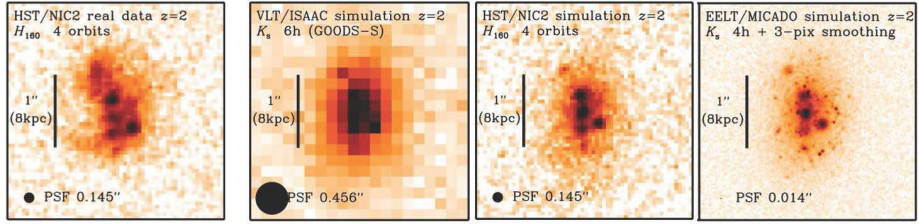}
\end{center}
\caption{Left: observation with HST of a massive star forming galaxy at $z \sim 2$. Right 3 panels: simulation of a comparable galaxy at $z \sim 2$ that is observed respectively from the ground with an 8-m telescope (seeing limited), with HST, and with MICADO on the E-ELT. Incredible detail on sub-100\,pc scales will be detectable, enabling one to probe the number, distribution, size, and luminosity of star-forming clumps in such galaxies. Such data will lead to major advances in our understanding of galaxy structure and evolution in the early universe and at the peak epoch of galaxy mass assembly.}
\label{fig:highz}
\end{figure}

Galaxy evolution and formation is the primary science driver at high redshift. Continuum and emission line mapping of high redshift galaxies will enable us to address questions concerning their assembly, and subsequent evolution in terms of mergers, internal secular instabilities and bulge growth. The resolution of better than 100\,pc, equivalent to 1$^{\prime\prime}$ imaging of Virgo Cluster galaxies, will resolve the individual star-forming complexes and clusters, which is the key to understanding the processes that drive their evolution.

\subsection{Black holes in Galaxy Centres}
The Galactic Centre is a unique laboratory for exploring strong gravity around the closest massive black hole. The fundamental goal is to measure the gravitational potential in the relativistic regime very close to the central black hole via stellar motions, using very faint stars that are likely present but cannot be detected nor studied with any other facility prior to the E-ELT. These motions may also reveal the theoretically predicted extended mass distribution from stellar black holes that should dominate the inner region, as well as test for a distributed component of dark matter.
Achieving the best astrometric precision -- also in the context of intermediate mass black holes in Galactic star clusters, ultra-compact dwarf galaxies, and dwarf spheroidals -- is the topic of several on-going studies that address the problem from different perspectives\cite{mas16,rod16}.
One of the outstanding questions in galaxy evolution concerns the cores and cusps in galactic nuclei, the role of central supermassive black holes, the mechanisms of mass transport into these central regions, and the influence of the galaxy-scale and larger environment. AGN feedback is also a crucial issue that appears to be responsible for quenching star formation and turning massive star-forming galaxies into passive spheroids. In the local Universe MICADO will fully exploit the much higher angular resolution of the E-ELT for studying massive black holes and the nuclear stars and gas at the low mass end, and at larger distances (and thus, effectively also at the higher mass end). Studies of large samples of QSOs at high redshift are complementary to the more detailed studies of the nuclei of nearer galaxies. At high-redshifts the resolution and sensitivity of MICADO will, for the first time be sufficient to spatially resolve the nuclear regions and study the growth of black holes and bulges in the epoch of galaxy formation.

\subsection{Resolved Stellar Populations}

\begin{figure}
\begin{center}
\includegraphics[width=17cm]{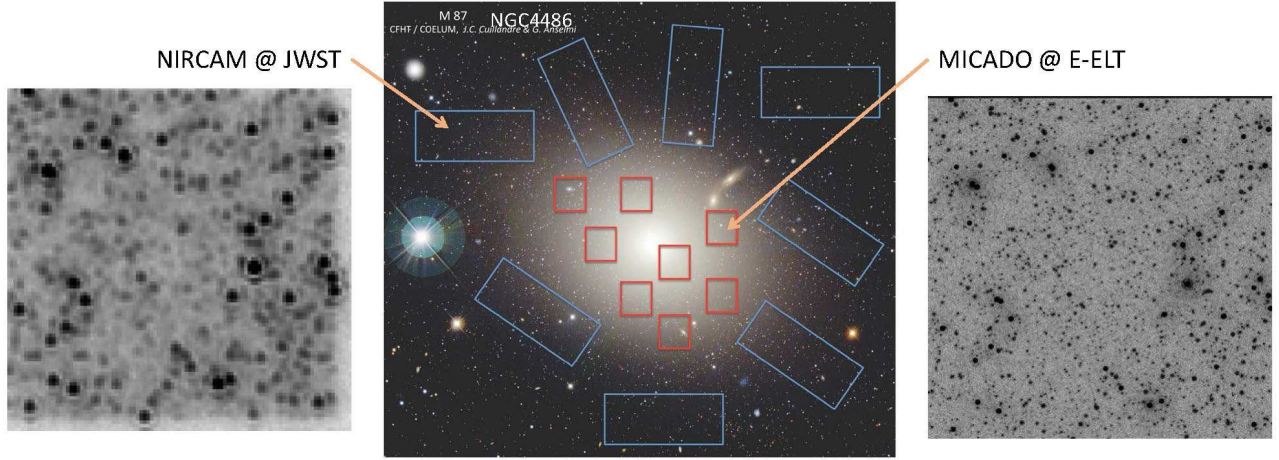}
\end{center}
\caption{Optical image of an elliptical galaxy in the Virgo cluster, with possible fields that could be surveyed by JWST (blue rectangles) or MICADO (red squares) overlaid. As an illustration of the power of spatial resolution in measuring resolved stellar populations, the same crowded field has been simulated for JWST (left) and MICADO (right). MICADO is expected to push 3\,mag deeper than JWST in such situations.}
\label{fig:stellarpops}
\end{figure}

Globular Clusters, among the oldest known components of the Milky Way, may also host intermediate mass black holes, the presence and masses of which would be derived from measurements of the cluster internal kinematics. Deriving the clusters' orbits would enable one to address questions about the formation and evolution of the Galaxy. The internal kinematics of Dwarf Spheroidal Galaxies will also shed light on this issue, by revealing the amount and distribution of dark matter in these objects; and hence testing models of structure formation, with respect to the clumps (sub-haloes) in which the dwarf spheroidals reside.
In addition to directly observing galaxies in the early universe, an alternative way to probe a galaxy's evolution is through its star formation history, which can be assessed using colour-magnitude diagrams that trace the fossil record of the star formation. Spatially resolving the stellar populations in this way is a crucial capability, since integrated luminosities are dominated by only the youngest and brightest population. MICADO will extend the sample volume from the Local Group out to the Virgo Cluster and, through its capability to perform photometry in crowded fields, push the analysis of the stellar populations deeper into the centres of these galaxies where JWST cannot\cite{dee11,gre12,sch14,gul14}.
The excellent performance of MICADO in crowded fields and its large field of view are also key to understanding star formation itself, whether in nearby galaxies or galactic star clusters. Deep multi-colour images will provide a detailed census of stellar types and ages in a variety of star forming environments, yielding important results about the form of the initial mass function.

\subsection{Coronagraphy}
One of the fundamental science drivers for the E-ELT is the detection and characterisation of extrasolar planets, ultimately rocky exoplanets in the habitable zone. MICADO will take the first steps along this road by providing a high contrast imaging capability that enables one to push in to very small angular separations\cite{bau14}. Specifically, it will provide access to 10--20\,AU orbits around young stars at distance of 100--150\,pc which are inaccessible to the current generation of exoplanet cameras such as SPHERE. In addition, it will also access 1--2\,AU orbits around nearby young stars, enabling the detection of less massive planets as well as the inner structure of circumstellar disks. Crucially, this range of orbits overlaps those detected with radial velocity techniques, opening the exciting opportunity to directly measure the luminosity of a planet whose mass is also known.

\subsection{Solar System}
At the other extreme of the distance scale from the early universe, but also related to planetary science, are studies of solar system objects. MICADO's large field of view is well matched to the angular sizes of planets such as Venus, Jupiter, and Saturn. Time resolved observations at high spatial resolution are an important part of understanding their atmospheres and weather systems.

\subsection{Time Resolved Astronomy}
One of the frontiers for observational astrophysics is the temporal domain below 1\,sec, where the physics of compact objects becomes important.
While radio, X-ray, and $\gamma$-ray astronomy routinely observe at much shorter timescales, leading to a number of fundamental discoveries, at optical and infrared wavelengths accessibility to this regime has been limited by sensitivity.
Neutron stars are testbeds for extreme phenomena, ranging from strong gravity to the post-supernova evolutionary path of massive stars. One class, magnetars, are observed as soft $\gamma$-ray repeaters or anomalous X-ray pulsars. More than half of the currently known magnetars are expected to be detectable by MICADO, and tracing their near-infrared light curves provides key constraints on the physical models.


\section{The MICADO-MAORY System}
\label{sec:micadomaory}

The MICADO instrument and MAORY adaptive optics system are being developed by different consortia.
In order to achieve a seamless integration, and to ensure they operate as a single system at the telescope, defining the global architecture and interfaces are key steps.
Following the decision by ESO that the MICADO-MAORY system should use the `straight-through' port of the E-ELT's pre-focal station and that MAORY should provide a second instrument port, the most critical aspects are the optical interface, structural architecture, and mechanical interface. Together these define the MICADO-MAORY system performance, while the mechanical interface also drives the accessibility of MICADO for integration and maintenance, as well as the level to which the development of the project designs can proceed independently.

\begin{figure}
\begin{center}
\includegraphics[width=17cm]{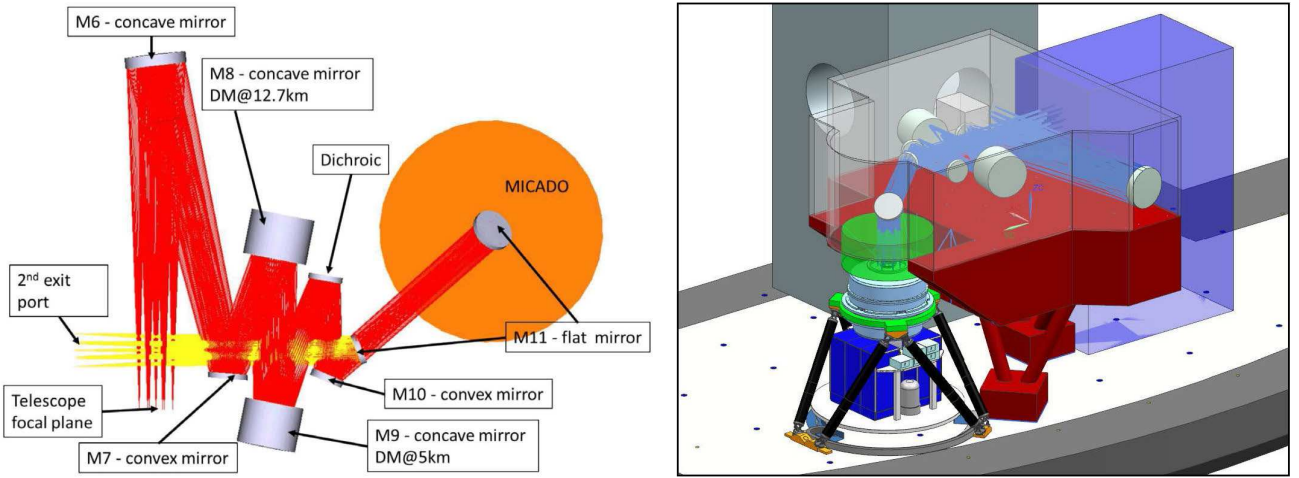}
\end{center}
\caption{left: baseline design of the MAORY optical relay, selected as a trade-off in terms of optical performance of the MICADO-MAORY system, and showing the location of MICADO. It was developed by the MAORY consortium and is described in detail by Lombini et al.(2016)\cite{lom16}. The dichroic here reflects optical and infrared light, and transmits only 589\,nm wavelengths (the beam path to the LGS WFS is not shown). Two mirrors are labelled as M11: the first is switchable, and used only if the light should be reflected to the 2nd instrument port; the default is for the light to be reflected down by the other M11 flat fold mirror into MICADO.
Right: the baseline mechanical concept is that MICADO is supported independently, and can be moved into a recess in the MAORY bench underneath the last fold mirror of the optical relay. This allows excellent accessibility for installation, testing, and maintenance. And the support structures for MICADO and MAORY can be developed relatively independently, vastly simplifying the whole interface issue. The purple box represents the volume reserved for a possible second client instrument for MAORY.} 
\label{fig:maoryrelay}
\end{figure}

In the context of the optical design, several options for the MAORY optical relay were compared in terms of wavefront error, geometric distortion, focal ratio, field curvature, exit pupil location, and focal plan location.
These criteria were assessed together rather than separately due to their inter-dependencies. While having fewer optical elements is generally preferred, one aim was to keep the optical interface similar to that of the telescope. And important considerations were whether a powered entrance window for MICADO would be needed to compensate for field curvature and the resulting impact on the polychromatic strehl due to chromatic dispersion, and the vignetting at the cold pupil for the various optical interfaces.
The baseline design for the optical relay is shown in the left panel of Fig.~\ref{fig:maoryrelay} and described in detail elsewhere\cite{lom16}.

The mechanical interface was, at the most basic level, a choice between hanging MICADO underneath the MAORY bench (as adopted during the Phase~A study\cite{dav10}) and mounting it next to the MAORY bench.
A preliminary choice has to be made now in order to enable the design to mature to a sufficient level of detail for the next review milestones.
But it is also a time when there is incomplete information about the Nasmyth platform (location and relative motion of the attachment points, for example during tracking, and transmission of vibrations) and when a detailed study of vibrations and flexure of the instrument support structures has not yet been completed. As such, a viable back-up option is also being considered.

Mounting MICADO underneath the MAORY bench may lead to better integrated stability of the system.
However, the space available in all directions is limited, with severe implications on the arrangement of the co-rotating electronics. 
And accessibility is compromised both in terms of access during testing and commissioning phases as well as installation and maintenance, by the need to manoeuvre MICADO between the legs of the MAORY structure while avoiding the design volumes of other instruments on the Nasmyth platform.
In addition, this option would mean a significant additional effort would be needed to prepare for a `stand-alone' phase, because a completely different support structure, de-rotator, co-rotator, and cable-wrap would be required.

The preferred alternative is to fit MICADO into a recess in the MAORY bench underneath the last folding mirror. In this option, the entire interfacing issues is very much simpler, both in terms of design and access. In this concept, MICADO is supported and de-rotated on a separate structure, with the co-rotating electronics located between its legs. And the natural guide star wavefront sensor (NGS WFS) module\cite{spa16}, in which the SCAO sub-system is also located, is mounted rigidly to the top of the cryostat. During operation, the structure is fixed to the Nasmyth platform. But during installation and maintenance procedures, one can release the attachments, and in principle move the entire MICADO instrument and support structure on rails or rollers away from the MAORY bench, so that it can easily be accessed by the dome crane or other handling tools. This baseline option for the mechanical concept is shown in the right panel of Fig.~\ref{fig:maoryrelay} and described in detail elsewhere\cite{dio16}.

\section{MICADO Overview}
\label{sec:overview}

\subsection{Support Structure, De-rotation, and Cable-wrap}

\begin{figure}
\begin{center}
\includegraphics[width=17cm]{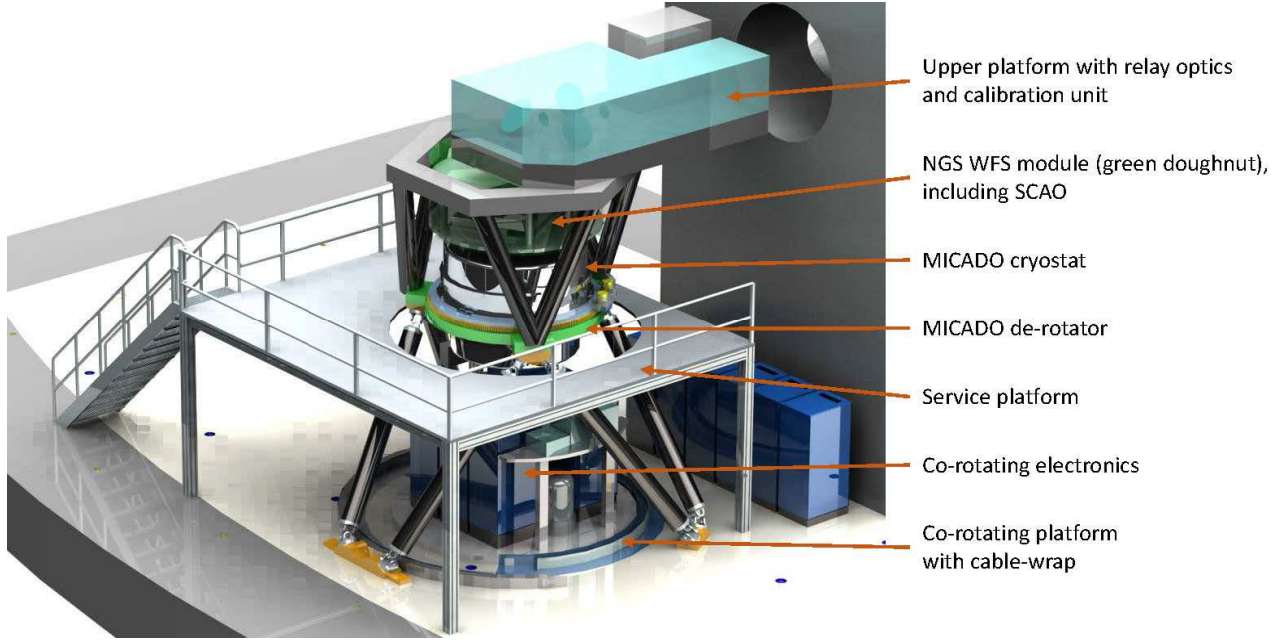}
\end{center}
\caption{Hexapod concept for supporting the de-rotator and MICADO cryostat and with the entrance focal plane at a height of 4.2\,m above the Nasmyth platform. Below the cryostat are the electronics on the co-rotating platform, which also houses the cable-wrap. Mounted on top of the cryostat is the NGS WFS module (for both MCAO and SCAO). In this view, the instrument is shown in the 'stand-alone' deployment configuration, which requires an additional small platform above, supported on a second hexapod, to hold the MICADO calibration assembly as well as the relay optics that will be used during the MICADO AIV phase.}
\label{fig:support}
\end{figure}

The horizontal optical axis from the telescope is 6\,m above the Nasmyth platform, and the focus is only 1\,m from the pre-focal station.
The design of MICADO requires that the focal plane is re-imaged at a height of 4.2\,m with the optical axis vertically oriented.
The rationale is to allow space above the entrance window of the cryostat for the mirrors of the relay optics (including folding the optical path downwards) and also the NGS WFS module; and to leave sufficient space for the electronics to be mounted on a co-rotating platform between the legs of the support structure below the cryostat (to keep cabling short where required). The co-rotating platform also houses the cable wrap to allow connections to the observatory services (cooling, power, signals, etc).

The design and analysis of the support structure\cite{nic16} and de-rotator\cite{bar16} are described elsewhere. Here it suffices to say that the structure is based on a hexapod that supports the cryostat at its centre-of-gravity to minimise the impact of residual displacements and dynamics of the camera mechanics.
It is likely that the lower end of the legs will have to be mounted to a rigid frame rather than directly attaching to the Nasmyth platform - both because of the location of the attachment points and to enable moving the entire structure away from the MAORY bench during maintenance.
The upper end of the legs support the de-rotator ring from below.
Field de-rotation is performed by rotating the whole cryostat rather than using a K-mirror due to the large ($>70^{\prime\prime}$ diagonal) field of view. It must be performed to a precision of order 10$^{\prime\prime}$, corresponding to about 1/3 of a pixel at the edge of the field, and move a mass in excess of 4000\,kg (which includes the cryostat and the NGS WFS module). This imposes strict requirements on the de-rotator rigidity, mechanism precision, and bearings.
In order to keep cable lengths short -- a requirement for the detector controllers, and preferable for many other devices -- most of the electronics will be housed in cabinets that are mounted on a co-rotating platform underneath the cryostat. The exact arrangement of the cabinets, particularly with respect to the support structure hexapod, is under development.

\subsection{Cryostat and Internal Structure}

\begin{figure}
\begin{center}
\includegraphics[width=16cm]{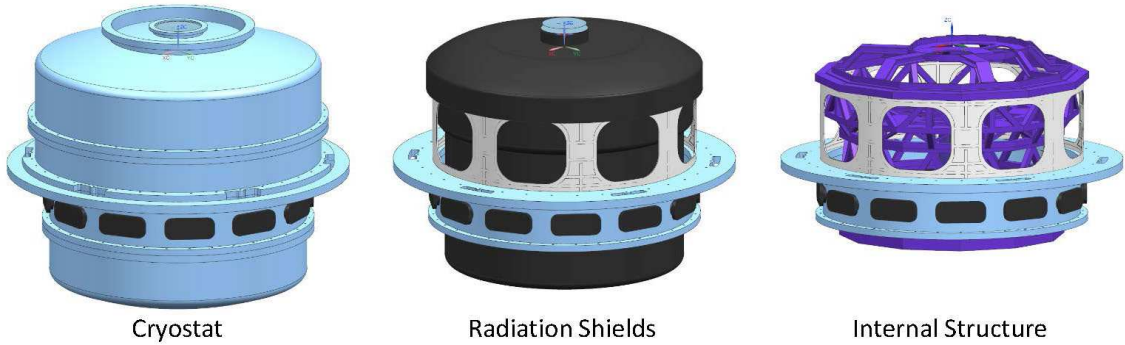}
\end{center}
\caption{Views of the cryostat for MICADO, which has a diameter of 2.1\,m. Left: external view (blue). The feedthroughs are below the central support ring; the NGS WFS module is mounted rigidly to the top of the cryostat. Centre: radiation shields (black), which are at $\sim140$\,K. Right: internal structure (purple), at 80--100\,K.}
\label{fig:cryostat}
\end{figure}

The arrangement of the mechanisms and folding of the optical path in the cryostat have led to a very compact design, with the diameter of the cryogenic internal structure being only 1.7\,m.
The cryostat is cooled with liquid nitrogen (LN2), either using a bath inside the instrument or continuous flow system with an auxiliary tank on the co-rotating platform to avoid putting the LN2 pipes through a cable-wrap.
Because the internal structure is operated at 80--100\,K the choice of materials for the internal structure and mirrors -- which have sizes up to 32\,cm -- is critical.
A typical choice for the cryogenic internal structure is aluminium.
And zerodur is an obvious choice for manufacturing the highest quality mirrors. But, because of the differing coefficient of thermal expansion (CTE) between the materials, the alignment at room temperature would be lost when the instrument is cooled to cryogenic temperatures -- requiring a lengthy series of complex alignment and cooling cycles.
It is possible to manufacture aluminium mirrors of the size needed. But with aluminium one cannot reach the required surface form and micro-roughness, so it needs to be covered with a thick $\sim200$\,$\mu$m layer of electroless nickel (NiP) that is polished before applying the reflective gold coating. The CTE mismatch between the Al mirror and the thick NiP layer is simply too high, inducing huge stress and leading to unacceptable warping that cannot be compensated.
An alternative is to use new materials such as aluminium silicon alloy -- with 42\% silicon the CTE matches that of NiP\cite{kin14}. The novel manufacturing technique of rapid solidification\cite{als12} has been used to produce this and other aluminium alloys. And there are claims that in some of them a lower micro-roughness can be achieved, meaning that a mirror made of such a material can be polished directly\cite{rsp}.
Currently, rapidly solidified aluminium alloys can be manufactured in sizes up to 1\,m, so one can consider using these materials for the entire internal structure of the cryostat.
The trade-off study for the mirrors and internal structure has not been completed, and so the final choice of material has not yet been made.

\subsection{Optical Concept}
\label{sec:opticalconcept}

\begin{figure}
\begin{center}
\includegraphics[width=16cm]{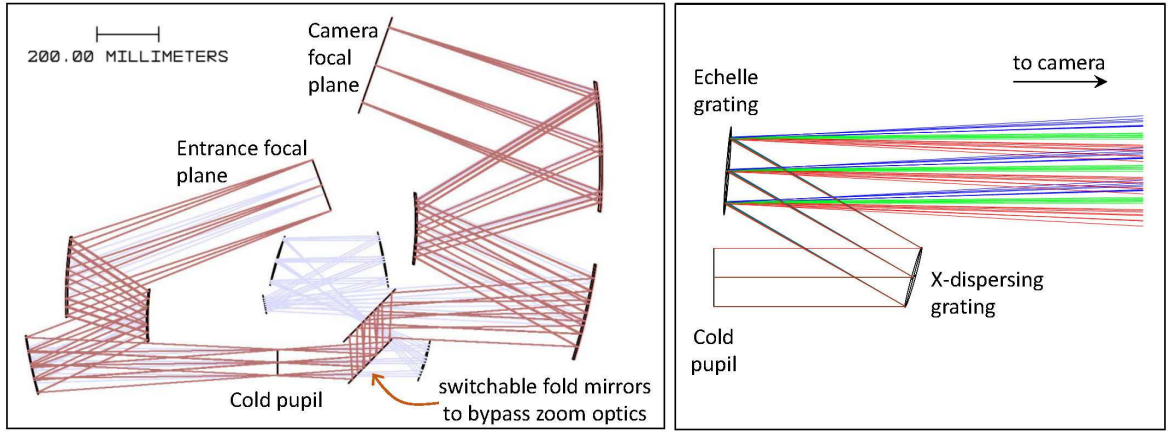}
\end{center}
\caption{left: The collimator and camera design for (i) the nominal 4\,mas pixel scale and $50^{\prime\prime}$ field of view (red) which includes two flat fold mirrors after the cold pupil, and (ii) the zoom option giving 1.5\,mas over a $\sim20^{\prime\prime}$ field (grey), which includes 4 additional working mirrors instead of the flat folds. Right: concept for the cross-dispersed grating spectrometer, the gratings for which can be moved into the beam path and which makes use of the same collimator and camera.}
\label{fig:opticalconcept}
\end{figure}

The rationale for the optical concept of MICADO is that there should be at least one optical path for imaging in which all the mirrors are fixed. This is important for astrometry and, together with the requirement that the gravity vectors of the mirrors do not change (i.e. gravity invariant rotation), ensures that geometric distortions in the instrument are as stable as possible for timescales of years.
The current design has retained the idea from the Phase~A study of using a 3-mirror anastigmat for both the collimator and camera, but has developed it significantly. One of the key improvements is that the zoom pixel scale and spectroscopic capability are now incorporated into the main optics and exploit the full focal plane array (Fig.~\ref{fig:opticalconcept}).
In this concept, the zoom pixel scale (1.5\,mas over a $\sim20^{\prime\prime}$ field rather than 4\,mas over $\sim50^{\prime\prime}$) is enabled by including four additional mirrors in the beam path.
The collimator and camera mirrors remain in exactly the same positions for both configurations; and the zoom optics are also fixed.
To switch from the smaller to the larger pixel scale, one simply bypasses the zoom optics by moving the two flat fold mirrors in.
Thus the former provides a configuration in which all the mirrors in the beam path are fixed; the only movable mirrors in the latter are flat folds.
In both cases, the resulting design is highly stable.

The spectrometer is designed for compact sources, with the aim of providing high spectral resolution ($R\simgt10000$) with a wide wavelength coverage.
Aditionally, in contrast to the Phase~A design where the spectroscopic module was in an auxiliary arm that fed a single extra detector, a new requirement was to make use of the same optical path and large detector mosaic as the imaging mode. 
A variety of design options using prisms and gratings were assessed.
The selected concept uses two gratings, one as an echelle working in 3rd to 8th order and one as a cross-disperser working in 1st and 2nd order, which are moved in after the cold pupil, so that the beam bypasses the zoom optics.
There is a single fixed configuration for the spectrometer, and the selection of either 0.8--1.35\,$\mu$m IJ bandpass or 1.45-2.4\,$\mu$m HK bandpass is done solely with order sorting filters.
Because of the location of the spectral traces on the focal plane, the slit length is limited to a 3$^{\prime\prime}$.
This is long enough to nod a point source back and forth along the slit in order to simultaneously obtain the sky background.
A restriction on the usage arises because the atmospheric dispersion compensation is done only at the cold pupil rather than before the focal plane mask, and means that the slit should typically be oriented along the parallactic angle.
This is not a concern when obtaining spectra of unresolved sources.

High contrast imaging is achieved within the scope of the imaging optics described above, by inserting appropriate masks or stops in the entrance focal plane and the cold pupil. It thus does not impact the optical concept.
High time resolution measurements also do not require additional optics, but drive instead the detector control software, observing preparation software, pipeline software, and operational scenarios.

\subsection{Cold Optics Instrument}

\begin{figure}
\begin{center}
\includegraphics[width=9cm]{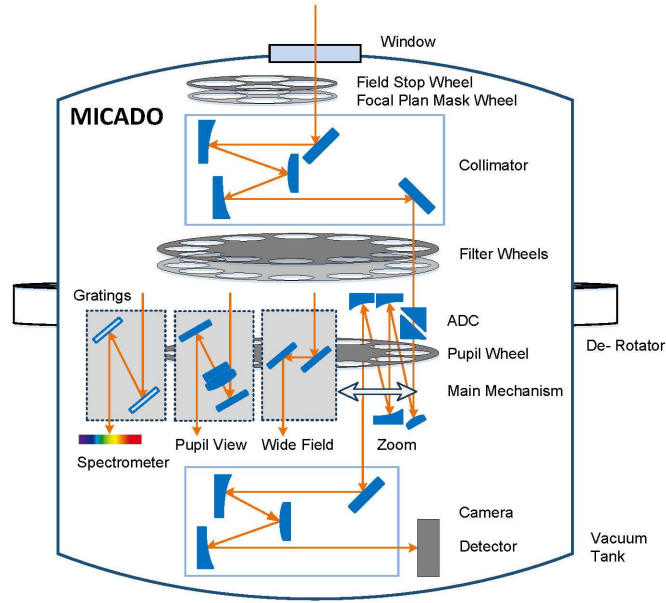}
\end{center}
\caption{Functional diagram of the MICADO cold instrument, summarising the optical layout and the mechanisms in the cryostat. Functional diagrams for the NGS WFS module (SCAO and its calibration unit), and the MICADO calibration unit are not included.}
\label{fig:functional}
\end{figure}

\begin{figure}
\begin{center}
\includegraphics[width=17cm]{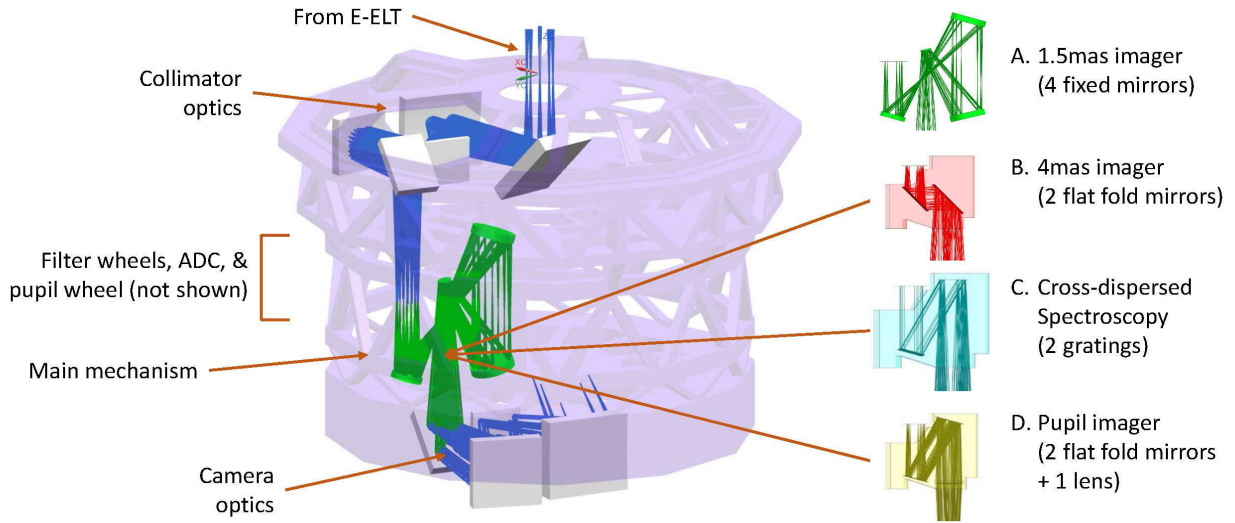}
\end{center}
\caption{Overview of the main optical path as it is folded into the cryostat. The collimator and camera are oriented horizontally at the top and bottom of the internal structure. Between them are most of the movable components: the filter wheels, ADC, and pupil wheel are not shown. This figure illustrates how the main mechanism enables one to switch between option (A) 1.5\,mas imaging, which uses fixed mirrors mounted to the internal structure and separate from the mechanism, and any of (B) 2 flat fold mirrors for 4\,mas imaging, (C) 2 gratings for cross-dispersed spectroscopty, or (D) 2 flat fold mirrors and a lens for pupil imaging.}
\label{fig:optics}
\end{figure}

\begin{figure}
\begin{center}
\includegraphics[width=16cm]{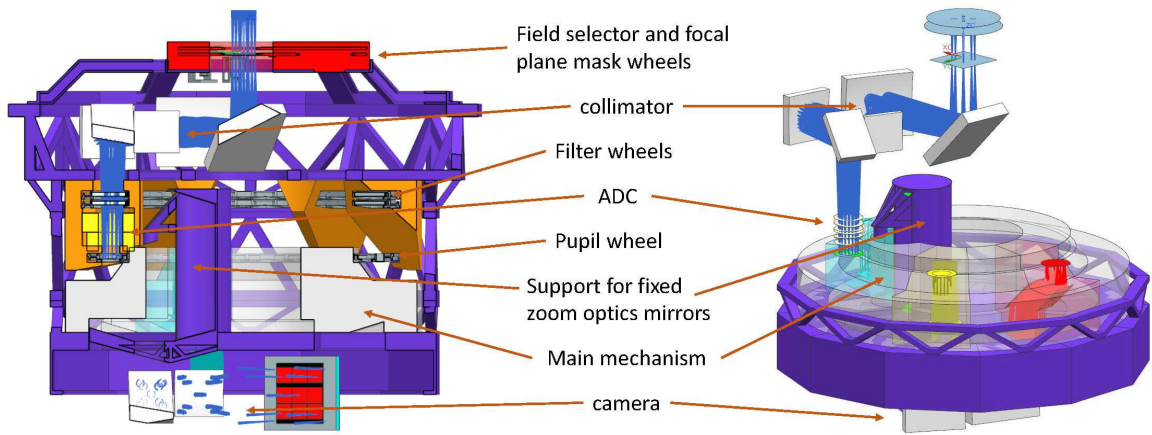}
\end{center}
\caption{Overview of the mechanisms inside the cryostat. At the top in red are the field selector and focal plane mask wheels. The filter wheels, ADC, and pupil wheel are mounted together in the orange structures, and comprise a single deliverable sub-system. Below these, the main mechanism in pale grey, which moves between the upper and lower mounts for the fixed zoom optics mirrors, is another deliverable sub-system. The cold internal structure is drawn in purple.}
\label{fig:mechanisms}
\end{figure}

The cold instrument design permits several sub-systems (each comprising some of the components described below) to be defined in a relatively simple way. These can be designed, assembled and tested independently by different partners in the consortium, and then efficiently integrated together.
The description here focusses on the various components, in order along the beam path. The functional diagram of the cold instrument shown in Fig.~\ref{fig:functional} summarises both the main optical path shown in Fig.~\ref{fig:optics} and the mechanisms shown in Fig.~\ref{fig:mechanisms}.

After the entrance window, the first element is the mask wheel at the entrance focal plane. This holds masks for the imaging fields, slits for spectroscopy, several coronagraphs for high contrast imaging, and also a point source mask for mapping instrument distortions. In order to pack these components closer together (e.g. the coronagraphs, which each transmit only a 6$^{\prime\prime}$ field, may all be within a 50$^{\prime\prime}$ envelope on the wheel) there is a second masking wheel next to it that restricts the transmitted field appropriately.

There are fold mirrors both before and after the collimator, which is mounted horizontally on an optical bench. It is followed by several key mechanisms.
In order to fulfil the requirement that there are at least 30 imaging filters available, the design incorporates two filter wheels that pass close to the pupil. The beam size here is nearly 110\,mm, so the wheels are large. Each contains 18 slots, which allows also for blocking options, spectroscopic order sorting filters, a neutral density filter, and a few spare positions. They are supported and driven around the edge, rather than centrally. As such, they do not impinge on the core of the instrument, which is available to provide structural support, especially for the fixed zoom optics.

After the filter wheels is the atmospheric dispersion compensator (ADC). This is required to obtain point-like sources for imaging, to enable astrometry, and to maximise sensitivity. Although it adds 8 surfaces to the optical path, the loss of throughput is more than compensated by the reduction in chromatic dispersion for zenith angles $>35\deg$ in K-band, $>10\deg$ in H-band, and $>5\deg$ in J-band. As such, there is no need to consider whether the device should be removable. 
The location and performance of the ADC has been the subject of a major study. One of the conclusions is that because there is no accessible pupil in the MAORY optical relay (and the pupil would anyway be unacceptably large for such a device), and because of the difficulties associated with locating ADCs close to a focal plane, the ADC in MICADO will consist of counter-rotating paired wedges and is located next to the cold pupil, which has an 80\,mm diameter.
Describing this device in detail -- manufacturing tolerances, control scheme and precision, performance, distortion -- is beyond the scope of this overview.

Immediately below the ADC is the pupil wheel, the design of which is similar to the filter wheels. It contains various cold stops as well as masks for pupil plane coronagrpahy, sparse aperture masking\cite{lac14}, and for calibrating non-common path aberrations. Despite the large size of the wheel, the only part requiring high precision is where it intersects the instrument pupil; the other support bearings have less tight positioning requirements, simplifying the design of this component.

The main mechanism holds the optics modules for the various observing modes. The empty position enables imaging at the smaller pixel scale. The other positions move in either two flat fold mirrors to switch to the larger pixel scale, gratings for spectroscopy, or two flat fold mirrors and a lens for pupil imaging.

A final fold mirror that directs the light into the camera which, like the collimator, is mounted horizontally on an optical bench. The last unit is the focal plane array, which holds the mosaic of $3\times3$ 4k$^2$ detectors. Currently we include the ability to cool these to 40\,K with a pulse tube cooler, since experience with other near-infrared instruments indicates that some aspects of their performance may be considerably better at that temperature.

\subsection{Control and Pipeline Software}

Software plays an increasingly important role in instrument projects.
In particular, adaptive optics instruments have complex operational set-ups and data processing requirements. The scope of the software extends far beyond controlling mechanisms and performing the observations; it is mandatory to include provision of tools to assist in observation preparation, on-line data quality and AO performance assessments, data processing pipelines, and PSF reconstruction to assist in data analysis.

MICADO will make use of standard tools similar to those used for VLT instruments.
Going beyond the standard exposure time calculator that is used for proposal preparation, SimCADO\cite{les16} is a instrument simulation tool that allows the user to evaluate the observability of specific science targets and optimise the observation strategy in terms of signal-to-noise (particularly in complex cases such as time resolved imaging), astrometric performance, or other data quality metrics.
For observation preparation, the usual phase 2 preparation tool will be supported by a range of other tools.
Common tasks include selecting adaptive optics guide stars from the technical field around MICADO, ensuring that these remain accessible for the entire planned dither sequence, and supplying an estimated map of AO performance. In some cases, secondary guide stars (that provide additional feedback from the MICADO science detectors) can also be chosen.
For time resolved applications, it is necessary to define specific detector windows that can be read out fast, given that the achievable frame rate depends on the number and size of windows.
For spectroscopy, it may be necessary to include blind-offsets;
for solar system science, a link to external ephemerides;
and for high contrast imaging, support of pupil stabilisation.

At the other end of the observing flow, the pipeline\cite{kle15} has to support all observing modes: standard imaging, astrometric imaging, spectroscopy, time resolved imaging, and, to a lesser extent, high contrast imaging.
An important aspect of data processing for AO instruments is obtaining knowledge of the PSF, an issue that has often been omitted from the scope of instrument projects but the need for which has also been emphasized\cite{dav12}.
A significant effort is being put into this aspect for MICADO, covering a variety of aspects.
These include PSF reconstruction from SCAO and MCAO\cite{wag16}, both in the high and low signal-to-noise regimes, adaption to the specifics of the E-ELT, the use of calibration stars and non-common path aberration measurements to help guide the reconstruction, and also assessing whether it is better to use occasional bursts of the full data from the real-time computer or a continuous sub-sampling of available data.
Related to this is a parallel development of myopic deconvolution optimised for a spatially and temporally variable PSF which may enhance the final analysis of the data, particularly for crowded stellar fields.

\section{Conclusion}
\label{sec:conc}

MICADO is the first light imaging camera for the E-ELT. The consortium designing, building, and commissioning the instrument comprises institutes in Germany, France, the Netherlands, Austria, and Italy. The project entered Phase~B in September~2015 and the next milestone is consolidation of the basic design concept together with the system and sub-system specifications and interfaces, followed in 2018 by the preliminary design review.

The instrument is optimised to operate with the multi-conjugate laser guide star adaptive optics system MAORY. The two consortia are also jointly developing a simple and robust single-conjugate natural guide star adaptive optics system. The observing modes offered by MICADO include: standard imaging, astrometric imaging, high contrast imaging, time resolved imaging, and slit spectroscopy. Observers will be assisted with dedicated preparation tools, data processing pipelines, and PSF reconstruction.

\acknowledgments     

RD thanks Andy Shearer (National University of Ireland, galway) and Vik Dhillon (University of Sheffield) for their valuable contributions on high time resolution astronomy with MICADO.
USM and IAG acknowledge support by the BMBF Verbundforschung.



\end{document}